\documentstyle[twocolumn,prl,aps,epsf,floats]{revtex}
\input epsf
\epsfverbosetrue

\begin{document}

\twocolumn[\hsize\textwidth\columnwidth\hsize\csname %
@twocolumnfalse\endcsname

 
\preprint{Submitted to {\em Physical Review Letters} }

\title{Counterion Condensation in Strong, Flexible Polyelectrolytes}
\author{Mark J. Stevens and Steve Plimpton}
\address{P.O. Box 5800, MS 1111, Sandia National Laboratories, 
Albuquerque, NM 87185}
\date{November 19, 1996}

\protect\maketitle
\protect\widetext

\begin{abstract}
We present results of molecular dynamics simulations of strong, flexible
polyelectrolyte chains in solution with added salt. 
The effect of added salt on the polyelectrolyte chain structure 
is fully treated for the first time as a function of polymer density. 
Systems above and below the Manning condensation limit are studied.
The chain contraction due to added salt is weaker than expected
from simple screening arguments.
The chain structure is intimately tied to the
ion density near the chain even for chains in the counterion condensation
regime. 
In contradiction to Manning counterion condensation theory,
the ion density near the polymer chain depends on the amount of added
salt, and above the condensation limit the chains significantly contract 
due to added salt.

\end{abstract}
]

\narrowtext

Polyelectrolytes are a very important class of polymers, because they are
one set of water-soluble polymers.
They comprise a class of biopolymers, e.g. DNA, RNA, and polysaccharides.
Their water solubility makes polyelectrolytes an important
class of synthetic, commercial polymers used for example, as the key
water absorbing ingredient in disposable diapers.
Understanding the effect of added salt on polyelectrolyte systems is
a basic issue for polyelectrolytes.
The addition of salt is a key means to alter polyelectrolyte structure 
and properties, and distinguishes polyelectrolytes from neutral polymers.
Nature takes advantage of this in biological systems, and the influence of
salt is common in industrial applications.
However, our understanding of polyelectrolyte structure is limited,
and consequently, so is our understanding of polyelectrolyte system properties
\cite{Schmitz94,Foerster95,Stevens95,Barrat96,Micka96}.
Recently, simulations have been able to determine the structure of
strong, flexible polyelectrolytes in salt-free solution \cite{Stevens95}.
In this Letter, we present results of polyelectrolyte simulations with
added salt. 
In particular we examine the influence of the ionic distribution on the 
chain structure.

Our understanding of polyelectrolytes is poor because of the 
difficulties these systems present to both experiment and theory.
Direct measurements of chain structure, particularly at dilute concentrations
have yet to be done.
Recent simulations of salt-free polyelectrolyte systems have overcome
some of the major theoretical difficulties \cite{Stevens95}.
The two main theories\cite{OSF} of polyelectrolyte structure have been shown
by recent simulations \cite{Stevens95,Micka96} to be incorrect for flexible 
polyelectrolytes.
Theoretical works have tended to neglect entropy, in part because they
focused on DNA which is intrinsically very stiff.
For stiff chains entropy is a small contribution to the free energy.
In contrast, for flexible polyelectrolytes, treating entropy along
with the Coulomb interactions is essential and has only been done properly
in simulation \cite{Stevens95,Micka96}, although recent self-consistent
field theory calculations that treat entropy are very promising \cite{Donley96}.

One of the major theoretical difficulties is the calculation of the
ionic density about the chain. 
All calculations of chain structure use the Debye-H\"uckel (DH)
approximation.
This approximation is a linearization of the Poisson-Boltzmann (PB) equation,
valid when the Coulomb interaction energy is much less than $k_BT$.
Yet, polyelectrolytes are often highly charged and the Coulomb energies
can be stronger than $k_BT$. 
The PB equation \cite{Guldbrand84,Kjellander84,Stevens96} 
is a mean field approximation that neglects ion correlations, and is valid 
when typical counterion separations are larger than the Bjerrum length,
$\lambda = e^2/\epsilon k_B T$, where $\epsilon$ is the dielectric constant
of the solution (water) and $e$ is the electron charge.
More sophisticated methods  
(e.g. hypernetted-chain (HNC) and local density functional theory), 
have been applied, but like the PB method 
only the problems with simple, fixed geometries can be solved
\cite{Stevens96,Bacquet84,Gueron80}.
In particular, since DNA has a large intrinsic persistence length of a few 
hundred \AA\, many calculations treating DNA have been performed for cylinders 
\cite{Bacquet84,Gueron80,Bratko82,Lebret84a}.

Even though it is known that the DH and 
PB approximations can fail, these two approximations 
pervade the field because of their simplicity.
Particularly in biophysics, counterion condensation (CC) theory 
\cite{Manning78} and the
PB theory \cite{Kotin62,Anderson82,Stigter95} are the two main theories. 
The basis of CC theory is the DH solution 
of a fixed, linear charge distribution (charged line, charged cylinder, or a
line of point charges) \cite{Manning78}.
In Manning's theory, CC occurs when $\xi \equiv \lambda/a \ge 1$, where
$a$ is the charge separation distance on the polyelectrolyte.
For $\xi > 1$, counterions condense onto the chain renormalizing the 
charge per unit length so that effective distance between charges 
is $a^*=\lambda$.
The condensed ions are contained in a cylindrical volume of
radius, $R_{CC}$ which is salt-independent.
There is much controversy over the success of CC theory 
\cite{Gueron80,Stigter95,Manning96}.
Solutions of the PB \cite{Gueron80} and HNC equations \cite{Bacquet84}
for charged cylinders contradict CC theory showing that the ion condensation 
is dependent on salt concentration.
This is consistent with the CC regime occuring 
where the DH approximation breaks down \cite{Stevens96}.
However, CC theory and experiment often agree \cite{Stigter95,Manning96}.
By simulation we can treat the Coulomb interactions without approximation
and resolve some of these issues.

Through molecular dynamics simulations,
one of us has successfully characterized strong, flexible polyelectrolytes
in salt-free solution \cite{Stevens95,Stevens93}.
The fundamental model of polyelectrolytes in solution consists of
chains of $N$ charged monomers with $N$ counterions per chain
and some added salt.
The chains have bond length $b$ and charge separation $a$ along the backbone.
The neutral solvent is treated as a continuous, dielectric background.
The system properties will depend on the polymer density, $\rho_p$,
or equivalently, the monomer density, $\rho_m = N\rho_p$, and the
added salt density, $\rho_s$.
Using MD we solve this fundamental problem exactly.
The results of ealier salt-free simulations reproduced experimental 
measurements of the osmotic pressure and the peak position in 
the inter-chain structure factor.
Here we treat the case of systems with added salt.

Here, we extend the previous work \cite{Stevens95} to include added salt
\cite{Stevens95}.
The salt ions are modeled the same as the counterions.
In fact, the salt ions oppositely charged to the monomers are identical
to the counterions.
All ions interact via a purely repulsive Lennard-Jones (RLJ) potential 
with the cutoff at $2^{1/6}\sigma$.
This gives the particles a diameter of about $\sigma$.
All ions including the monomers are monovalent in this work.
The number of added salt ions ranges from 0 to $8N_c$, where $N_c$ is
the number of counterions which is typically 256 or 512.

The polymers are modeled as freely-jointed bead-spring chains.
All beads are charged ($b=a$).
The bond potential is the standard FENE (nonlinear spring) potential with
spring constant $k = 7\epsilon/\sigma^2$, and maximum extent, $R_0 = 2\sigma$,
where here, as throughout the letter, Lennard-Jones units are
used \cite{Duenweg91}.
The average bond length is $b = 1.1 \sigma$.
Chains with $N= $ 32 and 64 monomers are treated. 
The number of chains is usually 8 with some simulations of up to 32
chains.

The Coulomb coupling strength is determined by the Bjerrum length.
One set of simulations were performed for the same parameter set as 
in the earlier salt-free work, $\lambda= 0.83\sigma$ \cite{Stevens95}.
In order to study the effects of counterion condensation, we also performed
simulations with $\lambda = 3.2\sigma > a$ which corresponds to a fully
charged chain when mapped to sodium polystyrene sulfonate \cite{Stevens95}.
For these simulations we used the particle-particle particle-mesh algorithm
to calculate the long range Coulomb interactions \cite{Hockney88,Pollock96}.
This algorithm's computation time scales roughly as $M\log M$, where $M$ is 
the number of charged particles.
Such scaling is essential to perform these calculations as 
added salt increases the number of charged particles significantly.
The simulations performed here involve up to 16 times the number
of charged particles as for the corresponding salt-free 
simulations \cite{Stevens95}.
Even with the almost linear scaling these simulations are at the edge of 
computational capability.

\protect\begin{figure}
\epsfysize=18pc
\begin{center}
\leavevmode
\epsffile{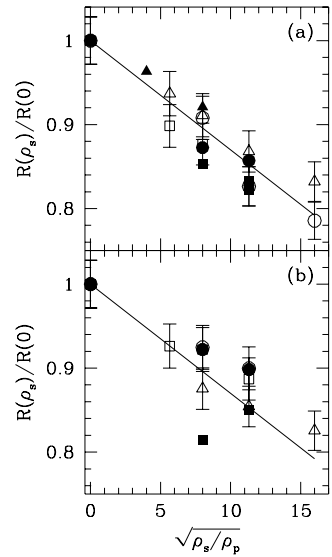}
\end{center}
\protect\caption{
The ratio, $B$, of the end-to-end distance, $R$, at salt density $\rho_s$
to $R$ at no salt for chain lengths $N=32$ (open points) and 64 (solid
points). Part (a) shows $\lambda=0.83\sigma$ and (b) has
$\lambda=3.2\sigma$.
The point types define the monomer densities: $\rho_m\sigma^3= 0.001
(\bigtriangleup )$, 0.01 ($\Box$) and 0.02 ($\bigcirc$).
The straight line is a guide to the eye and is identical in (a) and (b).
Error bars are only shown for $N=32$ to maintain clarity.
}
\label{fig:ratio}
\protect\end{figure}
\noindent

The temperature is maintained at 
$T=1.2\epsilon$, using the Langevin thermostat with damping constant 
$\Gamma=1\tau^{-1}$. 
The timestep is $0.015\tau$ \cite{Duenweg91}.
The number of timesteps for $N=32$ and $N=64$ is about $3\cdot 10^5$ and 
$8\cdot 10^5$, respectively.
This is sufficient to yield reasonable statistics.

The basic effect of the added salt is to screen the charged interactions and
contract the polyelectrolyte chain.
We quantified this contraction as a function $\rho_s$, $N$
and $\rho_m$ or $\rho_p$ by calculating 
the average, end-to-end distance, $R$.
We calculated the dimensionless ratio, $B = R(\rho_s)/R(0)$, which is the
ratio of $R$ at salt density $\rho_s$ to the salt-free value of $R$.
Simulations were performed at three monomer densities $\rho_m = $ 0.001, 0.01
and 0.02 $\sigma^{-3}$. 
For $N=32$ all these densities are below the overlap density.
For $N=64$, the salt-free overlap density is about $\rho_m=0.015\sigma^{-3}$ 
\cite{Stevens95}.

Figure \ref{fig:ratio} shows $B$ plotted against the ratio 
$\sqrt{\rho_s/\rho_p}$.
Plotting in this manner yields not only a rather linear dependence, but also
little if any dependence on $N$ or $\lambda$ within the
uncertainty of the data.
The square root dependence is related to Debye screening, but the ratio is not
between Debye lengths.
The numerator is related to the inverse Debye length for the added salt system,
$\Lambda_s = 1/\sqrt{8\pi \lambda \rho_s}$. 
However, the denominator is not the salt-free Debye length,
$\Lambda_0=1/\sqrt{4\pi \lambda \rho_m}$.
To obtain a quantity proportional to  $\Lambda_0/\Lambda_s$, we would need to
use $\rho_m$ in $\Lambda_0$ instead of $\rho_p$ ($\rho_m = N \rho_p$).
Using $\rho_m$ yields a small, but noticeable $N$-dependence.
We are skeptical that the apparent scaling in Fig. 1 will hold at larger $N$,
and view the data as the best approximate scaling.
But, as we will see, the screening length is different from these simple 
Debye lengths,
and recent calculations show the need of a wavelength dependent Debye length
\cite{Donley96}.
At large $\rho_s$ the scaling must break down, because
there will be a $N$-dependent crossover to a saturation region.

A 10\% reduction in $B$ occurs at about $\sqrt{\rho_s/\rho_p} = 8$ or
$\rho_s = 64 \rho_p$.
For $N=64$, this is where $\rho_s = \rho_m$; for $N=32$, $\rho_s=2\rho_m$.
To get an appreciable contraction beyond the zero salt state, one expects
that the screening due to the added salt must be stronger than the counterion
screening.
In other words, $\Lambda_s \approx \Lambda_0$.
Equality occurs at $\rho_s = \rho_m/2$ which is near where $B=0.9$.
Thus, when the added salt screens the monomer Coulomb repulsion about
equally to the counterion screening, contraction beyond the salt-free state
occurs.
One would like to understand the $N$ dependence better as well as the dependence
on $\rho_p$ that appears important, but simulations at larger $N$ are
presently too expensive.

At a sufficiently large $\rho_s$, the Coulomb interaction should be completely 
screened and the chain structure should be similar to a neutral chain.
In terms of Debye screening 
the monomer-monomer repulsion should be completely screened at $\Lambda_s = a$. 
If the monomer Coulomb interactions were treated at the DH level,
then at $\Lambda_s=a$ the Coulomb interaction becomes just a
short-ranged excluded volume interaction.
In this case, the chains would basically be neutral, 
and a further increase of $\rho_s$ would have negligible effect.
For $\lambda=0.83\sigma$, this occurs at $\rho_s = 0.04 \sigma^{-3}$.
At $N=32$, we have data spanning this regime in Fig. 1 at 
$\rho_m=0.02\sigma^{-3}$ for $\rho_s =$ 0, 2, 4 and $8\rho_m$.
No sign of a crossover to saturation appears.
At  $\rho_s = 8\rho_m$, $R= 10.8\sigma$. 
The value of $R$ for an equivalent neutral chain is 9.7$\sigma$.
We thus expect the saturation to start for this system at $\rho_s \agt 8\rho_m$
which is quite a bit larger than Debye screening suggests.
At $\rho_m=0.01\sigma^{-3}$ and $\rho_s=0.04\sigma^{-3}$, we find 
$R= 12.6\sigma$ which is far from the neutral value suggesting that $R$ 
will continue to decrease for larger $\rho_s$.
This result shows that screening due to explicit salt ions is weaker
than simple DH screening.
In simulations of salt-free solutions, we found that screening at dilute
densities was {\em stronger} when using explicit ions than 
for DH interactions \cite{Stevens96b}.
The results above show the opposite occurs for screening due to added salt;
the interactions between the added salt and the chains is screened by
the counterions stronger than DH predicts.

We now turn to results in the CC regime at $\lambda=3.2\sigma$.
In contradiction to CC theory, Fig. 1(b) shows that 
the chains contract similarly to the $\lambda=0.83\sigma$ case.
In CC theory adding more salt does not effect the environment near the chain. 
This suggests added salt should only have a weak effect on chain structure.
On the other hand, PB calculations have shown ionic density near the macroion
is dependent on salt density \cite{Bacquet84,Gueron80}.
Thus, our results agree better with PB predictions.

\begin{figure}
\epsfxsize=15pc
\begin{center}
\leavevmode
\epsffile{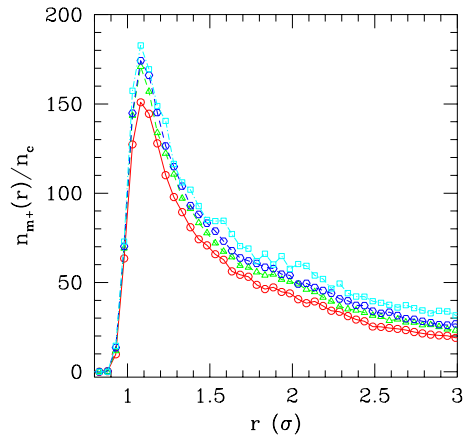}
\end{center}
\protect\caption{Ratio of the positive ion number density about a monomer
to the bulk counterion number density for $N=32$ at 
$\rho_m=0.001\sigma^{-3}$ and $\lambda=3.2\sigma$ for
$\rho_s/\rho_m=0$ ($\bigcirc$), 1 ($\bigtriangleup$), and 8 ($\Box$). 
This system is in the counterion condensation limit and has a dilute 
polymer concentration.
}
\label{fig:npc32}
\protect\end{figure}

We directly examine the ionic distributions in Fig. \ref{fig:npc32}.
The figure shows the normalized radial ion number density, $n_{\rm m+}(r)$, 
of all positive ions (i.e. counterion and positive salt) about a 
negatively charged monomer for $N=32$, $\lambda=3.2\sigma$ and 
$\rho_m=0.001\sigma^{-3}$ at several salt densities.
This distribution is calculated from the monomer-counterion radial
distribution function (rdf). 
Since the counterions and positive salt ions are identical particles,
their rdf's are identical (this was confirmed).
The number density is calculated by multiplying the rdf by
the number of positive ions (the chains are negatively charged).
We then normalized by the counterion density.

\protect
\noindent

The peak heights at different $\rho_s$ values are not identical.
For $\rho_s=\rho_c$, the peak is about 10\% higher than the salt-free case.
This is not a large difference in peak heights. 
For larger $\rho_s$ the relative peak height increase is smaller.
However, the peak position is not the only relevant spatial regime.
As $r$ increases to distances corresponding to just a few monomers 
away the distributions deviate as they must tend toward their asymptotic 
limit of $1+\rho_s/\rho_c$.
The limit is 9 for the largest $\rho_s$ in Fig. 2.
At $r=3\sigma$, there is a 50\% increase in the ion density for 
$\rho_s=8\rho_m$ over $\rho_s=0$.
Thus, {\em within the volume of the chain the ion density increases as
$\rho_s$ increases}.
Even for $\xi > 1$,
the repulsion between monomers distant along the chain is increasingly 
screened by increasing $\rho_s$, and thus the chains contract.
This contradicts CC theory which predicts no salt dependence of the ion
density near the chain.

\begin{figure}
\epsfxsize=15pc
\begin{center}
\leavevmode
\epsffile{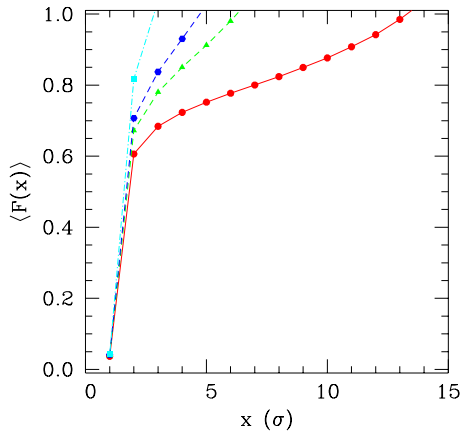}
\end{center}
\protect\caption{Fractional number of the positive ions within $x$ of a 
chain for the same systems as Fig. \protect\ref{fig:npc32}. 
$\rho_s/\rho_m=0$ (circle), 1 (triangle), 2 (hexagon) 
and 8 (square). 
The Manning condensation fraction for these systems is 0.65.
}
\label{fig:ccfig}
\protect\end{figure}

The number of ions near a chain can be directly counted. 
We define the distance between an ion and the chain to be the minimum
of the distances between the ion and each monomer in the chain. 
Using this metric, we can calculate the number of ions within a
distance $x$ to $x + dx$ of the chain.
Normalizing by $N$, we obtain the fractional number, $f(x)$.
Integrating $f(x)$ gives the total fractional number of ions within a
distance, $x$, which we define as $F(x)$.
If each chain on average has $N$ counterions within $x_N$ of it, then
$F(x_N) = 1$.
In the case of added salt, since the counterions and positive salt ions
are indistinguishable, we calculate $F$ where an ion is either a counterion
or a positive salt ion.
When $F(x) > 1$ for $\rho_s > 0$, this means that the number of
counterions and positive salt ions within $x$ of a chain is larger
than the number monomers on the chain.
This will clearly happen for large $\rho_s$.
Figure \ref{fig:ccfig} shows $F(x)$ for the same systems as 
Fig. \ref{fig:npc32}.
Because of steric repulsion, there are very few ions within $1\sigma$ of
the chains.
Within $2\sigma$ there is a substantial number of ions.
At $\rho_s = 0$, $F(2\sigma) = 0.60$ which is less than the Manning
fraction for linear chains, $1/\xi = 0.65$.
With increasing $\rho_s$, $F(2\sigma)$ increases to about 0.82 at
$\rho_s=8\rho_m$, well above the Manning fraction. 
Furthermore, within less than $3\sigma$, $F$ reaches 1! 
Thus, the situation is more complicated than CC theory predicts.

\protect
\noindent

All these results imply that extending counterion condensation theory to strong,
flexible polyelectrolytes fails to predict the correct salt dependence
of the chain structure.
The weakness of the Debye-H\"uckel approximation is especially
strong in the condensation regime and probably causes the CC predictions
to deviate numerically from ``exact'' values determined by simulation, but
it also is essential to treat the flexibility of the chain.
Given our strong disagreement with CC theory, why is CC theory
thought to be well confirmed by experiment?
First, counterion condensation in some sense clearly occurs near $\xi=1$
in the simulations \cite{Stevens95}.
It is the statistical physics of the counterion condensation which
we find to be different for predictions of Manning's CC theory.
Part of this disgreement occurs because for flexible chains
counterion condensation alters the chain structure, something
that beyond the original formulation of CC theory.
Measurement of the polyelectrolyte structure at dilute concentrations
has yet to be done, let alone a measurement of the
ion distributions near the chain where we find the major discrepancies.
Much of the agreement with experiment is based on the presence of
counterion condensation, not on the details of CC theory.
Furthermore, many experimental measurables depend only on the ion
distribution in the interstitial regime between macroions where
the DH approximation is often accurate \cite{Stevens96,Stigter95}.
These quantities are often not related to the chain structure or
the ionic distributions near the chain, and thus are
not good tests of CC theory's fundamentals.

This work was supported by the DOE under contract DE-AC04-94AL8500.
Sandia is a multiprogram laboratory operated by Sandia Corp., a
Lockheed Martin Company, for the DOE.




\end{document}